\documentclass[a4paper]{article}

\usepackage[english]{babel}
\usepackage[utf8]{inputenc}
\usepackage[T1]{fontenc}
\usepackage[colorinlistoftodos, color=green!40, prependcaption]{todonotes}

\usepackage{hyperref}
\hypersetup{colorlinks, allcolors=black}

\usepackage{subcaption}
\usepackage{graphicx}
\usepackage{ragged2e}
\usepackage{geometry}
\usepackage{sidecap}
\usepackage{amsmath}
\usepackage{bigints}
\usepackage{authblk}

\DeclareUnicodeCharacter{2212}{\ensuremath{-}}

\title{\huge Large area monocrystalline and surfactant-free \\
copper microflake synthesis}
\author[]{Elif Nur Dayi}
\author[]{Diotime Pellet}
\author[]{Priscila Vensaus}
\author[]{Fatemeh Kiani}
\author[]{Alan R. Bowman}
\author[]{Omer Can Karaman}
\author[]{Giulia Tagliabue\thanks{Correspondence email address:~giulia.tagliabue@epfl.ch}}
\affil[]{Laboratory of Nanoscience for Energy Technologies (LNET), STI, \\École Polytechnique Fédérale de Lausanne, 1015 Lausanne, Switzerland}

\begin{document}

\maketitle

\begin{abstract}

Copper is one of the most extensively studied materials for energy conversion and catalytic systems, with a wide range of other applications from nanophotonics to  biotechnology. However, existing synthesis methods are limited with many undesirable by-products and poorly defined morphologies. Here, we report a surfactant-free on-substrate wet synthesis approach that yields monocrystalline metallic Cu microflakes with (111) crystalline exposed surface. By systematically studying the growth mechanism, we achieve unprecedented sizes of more than 130 $\mu$m, which is two orders of magnitude larger than reported in most previous studies, along with higher aspect ratios of over 400.  Furthermore, we show distinctly higher stability against oxidation provided by the halide adlayer. Overall, our facile synthesis approach delivers an exciting venue for the emerging fields of catalysis and nanophotonics.

\end{abstract}


\section{Introduction} \label{sec:Intro}

The controlled growth of metallic nanomaterials with desired morphology, high purity, and well-defined crystallinity is crucial for many applications in plasmonics and nanophotonics, \cite{huang_freestanding_2011,xiong_kinetically_2005,gao_highly_2012}, optoelectronics\cite{karaman_ultrafast_2024,pan_large_2024}, flexible electronics\cite{sheng_copper_2022}, energy conversion and catalysis\cite{kiani_interfacial_2023,luc_two-dimensional_2019,huang_freestanding_2011,yang_synthesis_2017}. In particular, copper has emerged as a key material for energy applications: it is the most extensively studied catalyst for the CO$_{2}$ reduction reaction (CO$_{2}$RR), thanks to its unique ability to sustain the formation of carbon-carbon bonds on its surface and ultimately the generation of high energy-density multi-carbon products\cite{roberts_high_2015,kumar_photochemical_2012,zhao_charting_2022}. 

Catalytic, thermal and optoelectronic properties of Cu nanomaterials depend on the preparation technique and material quality\cite{dai_ultrastable_2017,li_thermal_2020,molahalli_properties_2024}. Although there is a plethora of methods for the growth of Cu nanoparticles in the literature, such as solvothermal\cite{wei_simple_2007}, microwave-assisted growth\cite{sreeju_microwave-assisted_2016}, electrochemical reduction\cite{zhang_electrochemical_2014} and wet synthesis\cite{zaza_increasing_2024,luc_two-dimensional_2019}; to date, there is no simple, efficient, and cost-effective method that allows controlled growth of large aspect ratio Cu nanomaterials with (111) exposed smooth surfaces. This specific surface orientation can be beneficial for a wide range of applications, for instance to steer the chemical selectivity of reactions\cite{luc_two-dimensional_2019,dai_ultrastable_2017} or to achieve unique plasmonic properties\cite{bowman_quantum-mechanical_2024}.

Table \ref{tab:lit-table} gives an overview of the state of the art in the synthesis of thin Cu nanomaterials with (111) orientation, namely nanoflakes, nanoplates and nanosheets. However, these methods offer either quite limited size, low yield or poor selectivity towards a specific geometry. To the best of our knowledge, most reported Cu flakes remain below a lateral size of only a few micrometres\cite{wang_sub-1_2023,luc_two-dimensional_2019}. In the few cases where the flakes are larger than >5 µm, they are not well separated from each other and are surrounded by undesired side products such as nanorods\cite{sheng_copper_2022,luc_two-dimensional_2019,tang_role_2017}. In addition, these methods utilize various surfactants and organic molecules including glucose, hexadecyltrimethylammonium bromide (CTAB), hexamethylenetetramine (HMTA); which can easily adhere to the surface after synthesis \cite{luc_two-dimensional_2019,fu_alternative_2020}. These organic surfactants can reduce the available active sites, restrict access to the surface or even interfere with chemical reactions and lead to misinterpretations of catalytic activity and selectivity\cite{li_surfactant_2012}. Removal of these molecules from the surface requires additional treatment, such as annealing or acid baths, which impairs surface quality. A surfactant-free method can be advantageous to avoid these challenges.  \\

\begin{table}[ht!]
\centering
\caption{Brief summary of reported syntheses for thin Cu materials with (111) surface orientation}
\label{tab:lit-table}
\begin{tabular}{|c|c|c|c|c|c|c|c|c|}
\hline
\textbf{Material} & \textbf{Method} & \textbf{Content} & \textbf{Lateral Size} & \shortstack{\textbf{Aspect}\\ \textbf{Ratio}} & \textbf{Reference} \\ \hline
Cu nanoplatelets & Electroreduction & \shortstack{\text{Cu(CN)}$_2^-$} & $\approx 200 \, \text{nm}$ & NA & \cite{henglein_formation_2000} \\ \hline
Cu nanoplatelets & Wet synthesis & \shortstack{\text{CuBr} \\ DPP \\ OLAM} & $\approx 1.019 \, \mu\text{m} \pm 0.519 \, \mu\text{m}$ & NA & \cite{zaza_increasing_2024} \\ \hline

Cu nanosheets & Wet synthesis & \shortstack{$\text{Cu(NO}_3\text{)}_2$ \\ L-Ascorbic Acid \\ HMTA \\ CTAB} & $\approx 1.7 \, \mu\text{m} \pm 0.5 \, \mu\text{m}$ & 340 & \cite{luc_two-dimensional_2019} \\ \hline

Cu nanosheets & Wet synthesis & \shortstack{$\text{Cu(NO}_3\text{)}_2$ \\ L-Ascorbic Acid \\ HMTA \\ TTAB} & $\approx 1.7 \, \mu\text{m} \pm 0.5 \, \mu\text{m}$ & 300 & \cite{fu_alternative_2020} \\ \hline

Cu nanoplates & Wet synthesis & \shortstack{\text{CuSO}$_4 \cdot 5\text{H}_2\text{O}$ \\L-Ascorbic Acid \\ CTAB} & $\approx 3.4 \, \mu\text{m}$& NA & \cite{wang_synthesis_2017} \\ \hline

Cu nanoplatelets & Wet synthesis & \shortstack{\text{CuCl}$_2 \cdot 2\text{H}_2\text{O}$ \\ D-Glucose \\ HDA \\ NaI} & $\approx 5 \, \mu\text{m}$& 300 & \cite{sheng_copper_2022} \\ \hline

Cu nanoplatelets & Wet synthesis & \shortstack{\text{CuBr}$_2$ \\ L-Ascorbic Acid \\ BPEI} & $\approx 8.03 \, \mu\text{m} \pm 3.18 \, \mu\text{m}$ & 22 & \cite{tang_role_2017} \\ \hline

Cu nanoplates & Wet synthesis & \shortstack{\text{CuBr}$_2$ \\ L-Ascorbic Acid \\ BPEI \\ Ag } & $\approx 10.97\, \mu\text{m} \pm 3.45 \, \mu\text{m}$ & NA & \cite{xu_preparation_2022} \\ \hline

Cu microflakes & \shortstack{Substrate-assisted \\ wet synthesis} & \shortstack{\text{CuSO}$_4 \cdot 5\text{H}_2\text{O}$ \\ L-Ascorbic Acid \\ KBr} & $>130 \, \mu\text{m}$ & 400 & \shortstack{This \\ work}\\ \hline

\end{tabular}
\end{table}

In this study, we present a straightforward route for surfactant-free growth of monocrystalline Cu microflakes and address the major bottlenecks in existing methods, such as low selectivity, reduced yield, and poorly defined morphologies.Our optimized recipe is constructed by carefully monitoring the effect of various parameters on the growth process, such as the reaction temperature, salt precursors, halide type and concentration and allow us to push the the lateral dimensions from just a few microns to over a hundred. We combine these efforts with advanced characterization techniques to establish the monocrystallinity, (111) orientation and the elemental composition of the flakes and gain insight to the growth mechanism. Remarkably, we show that thanks to the halide adlayer, the flakes exhibit significantly extended stability against surface oxidation, a phenomenon that challenges applications of Cu based nanomaterials. Lastly, while most flakes exhibit a smooth basal surface, specifically at high temperatures we notice flakes with distinct features such as steps, which can be of high interest for surface studies.These surfactant-free Cu microflakes with tunable morphology will enable studies of the optoelectronic properties of the monocrystalline metal as well as of energy and chemical conversion processes in hitherto unprecedented detail.

\section{Results} \label{sec:Results}

While synthesis of large area Au monocrystalline flakes is well established in the literature, their Cu counterparts have not been succesfully reported despite having high potential for applications in energy and photonics.  Here, we present a method for growing large-area Cu microflakes on a substrate. Adapting methods used for Au to Cu is not trivial, as the two materials differ significantly, e.g. in terms of oxidation behavior and lattice parameters. Thus, we systematically tested a range of chemical components and growth conditions to find a combination that promoted Cu flake growth.

Our synthesis configuration consists of a polypropylene tube that acts as a reactor, two borosilicate glass substrates placed inside the tube in a tilted orientation, a temperature controller and a heating plate, similar to the system previously reported \cite{kiani_high_2022}.  The reaction medium is ultrapure water (20 ml). To find the optimal precursor for flake growth, we tested several Cu salts with different anions such as $\text{SO}_4^{2-}$, $\text{NO}_3^-$, $\text{Cl}^-$ and $\text{CH}_3\text{COO}^-$, as the chemical affinities of the anions are known to be vital in shaping the crystallization and growth mechanisms \cite{chen_reaction_2014}.
We chose Cu sulfate pentahydrate ($\text{CuSO}_4 \cdot 5\text{H}_2\text{O}$, 10 mM) as Cu ion source, as it yielded fewer undesired by-products and the highest flake yield, as shown in Supplementary Figure 1. Since the formation of thin Cu flakes requires a slow reduction rate, we use L-ascorbic acid(30 mM) as a mild reducing agent \cite{xia_shape-controlled_2009}. Finally, we use potassium bromide (KBr, 4.2 mM)  as the capping agent, which we later discuss in detail. The reactor was maintained at 80°C in a light-tight water bath for 20 hours, which we refer to as the growth period (see Methods \ref{sec:Methods} for further details). This halide-assisted recipe was the first combination among many to consistently generate flakes of >10 $\mu$m grown on both sides of the substrates, thus we refer to it as the standard recipe and utilize it for parametric studies. After each synthesis, the flakes were imaged with an optical microscope and their size was analyzed using the commercial software Image J\cite{schneider_nih_2012}. The lateral size was measured by edge length for triangular flakes and by diagonal length for hexagonal and truncated flakes to ensure consistency with previous studies \cite{guo_facile_2006,li_edge_2012}. 

Nanoparticle formation can generally be divided into two phases: the initial nucleation phase, where individual atoms coalesce to form clusters or seeds, and the subsequent growth phase, during which these seeds develop into crystals with specific shapes. The growth of nanomaterials with a desired crystalline orientation can be achieved through the use of capping agents, whose binding affinity can promote or hinder the growth of a particular facet and control the overall morphology of the particles. Previous studies on halide-assisted growth of Cu and other metals have shown that Br$^-$ ions are selectively adsorbed at the \{111\} basal plane, which refers to the exposed surface, and allow surface passivation \cite{tang_role_2017,magnussen_-situ_1996,ghosh_many_2018, kiani_high_2022}. The here proposed halide-assisted wet-synthesis on glass substrates offers the following major advantages: The shaping and complexation effects of the halides can favor lateral anisotropic growth of the flakes, while the growth of nanoparticles on the substrate brings higher quality as aggregation, bending and contamination by small particles are avoided, in contrast to methods such as drop casting after synthesis.

\begin{figure} [ht!]
    \centering  
   \includegraphics[]
   {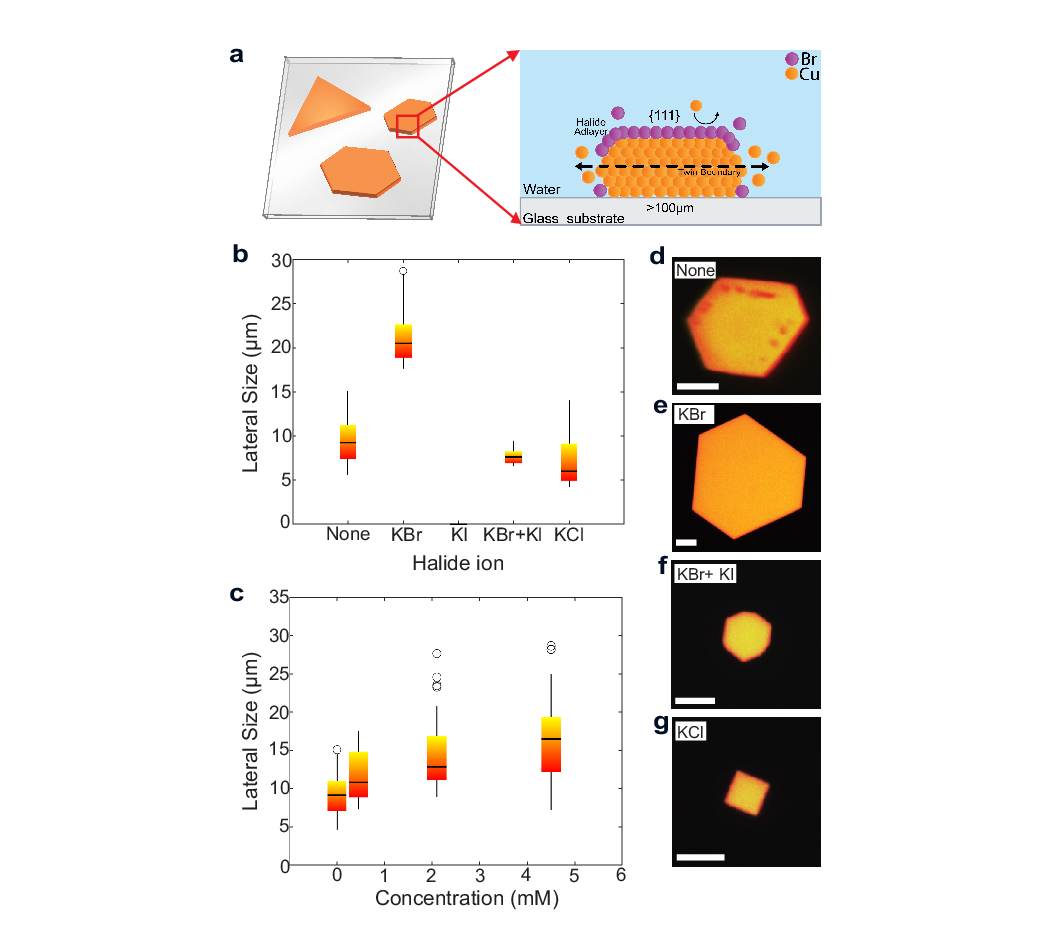}
    \caption{\justifying{\textbf{Overview of our growth of Cu microflake growth on substrate and the halide controlled morphology }
     \textbf{a} Illustration of the Cu flakes grown on substrate, planar (left) and cross-section (right) view. The halide adlayer (purple) promotes growth in the lateral direction. \textbf{b} Average lateral size obtained by using different halide ions. Resulting morphology for each condition are shown in c-f, with an exception for the use of KI, for which there was no visible formation of flakes. All scale bars indicate 5 $\mu$m. \textbf{c} Dependence of flake lateral size on KBr concentration. \textbf{d}  Anisotropic flakes with surface impurities were observed when no halide ions are used.  \textbf{e} Hexagonal flake obtained when using 4.2 mM of KBr.\textbf{f}  Etched flakes when KBr and KI are included simultaneously at equal concentrations of 4.2 mM.\textbf{g} Square-like flakes formed when 4.2 mM of KCl is used.  }}
    \label{fig-halide}
\end{figure}

Figure \ref{fig-halide} visualizes the Cu flake growth approach used in this paper and demonstrates the strong influence of halide ions on the control of Cu nanoflake lateral size and final morphology.  Figure \ref{fig-halide}b shows that the lateral dimension doubles when KBr is chosen as the halide ion compared to other halides such as KI and KCl or the case without halide. To better understand the exact role of KBr on the growth mechanism, we varied its concentration, as shown in Figure \ref{fig-halide}c. Up to 5mM, the average flake size is proportional to the KBr concentration. However, at more than 5 mM, we observed significant changes in flake morphology due to complexation and oxidative etching effects, which also increase with increasing concentration \cite{chen_high-yield_2014}. The effect of the high KBr content is visible on the scanning electron microscopy-energy dispersive X‐ray spectroscopy (SEM-EDS) elemental maps (see Supplementary Figure 2), where the flake becomes visibly thinner in the center. Since precise tuning of the KBr concentration is important to ensure a smooth surface while supporting lateral growth, we choose KBr as the halide of choice with the optimal KBr concentration of 4.2 mM. Apart from size, we also observe significant changes in the facets depending on the halide ion. For example, while it is possible to grow flakes up to 10$\mu$m without the halides, the facet growth is less controlled, resulting in a non-equilateral, elongated hexagonal shape, as shown in Figure \ref{fig-halide}d. When KBr is present as a shape-directing agent, we instead observe equilateral triangles and hexagons, a sign of isotropic lateral growth. In addition, flakes grown without halide tend to show dark spots on the surface on bright-light micrographs immediately after synthesis, which could be caused by partial oxidation. These dark spots are not observed in KBr assisted growth shown in Figure \ref{fig-halide}e, suggesting that the halide ions not only direct the shape but also act as a capping agent and protect the surface from oxidation, consistent with previous reports\cite{magnussen_-situ_1996}. No significant flake growth was observed when using KI and the combination of KBr with KI resulted in smaller flakes with a larger number of smaller side facets (Figure \ref{fig-halide}f), likely due to strong complexation effects. The flake size was further reduced when only KCl was used, with a significant change in surface orientation as shown in Figure \ref{fig-halide}f.

Upon investigating the role of halide ions, we employed numerous advanced characterization techniques evaluate the stability, surface quality, crystallinity and metallic nature of the flakes. Unless specified otherwise, all measurements were performed on Cu flake samples grown on a borosilicate glass substrate using the standard recipe with 4.2 mM KBr presented earlier in the text.

\begin{figure} [ht!]
    \centering  
\includegraphics[width=17cm]{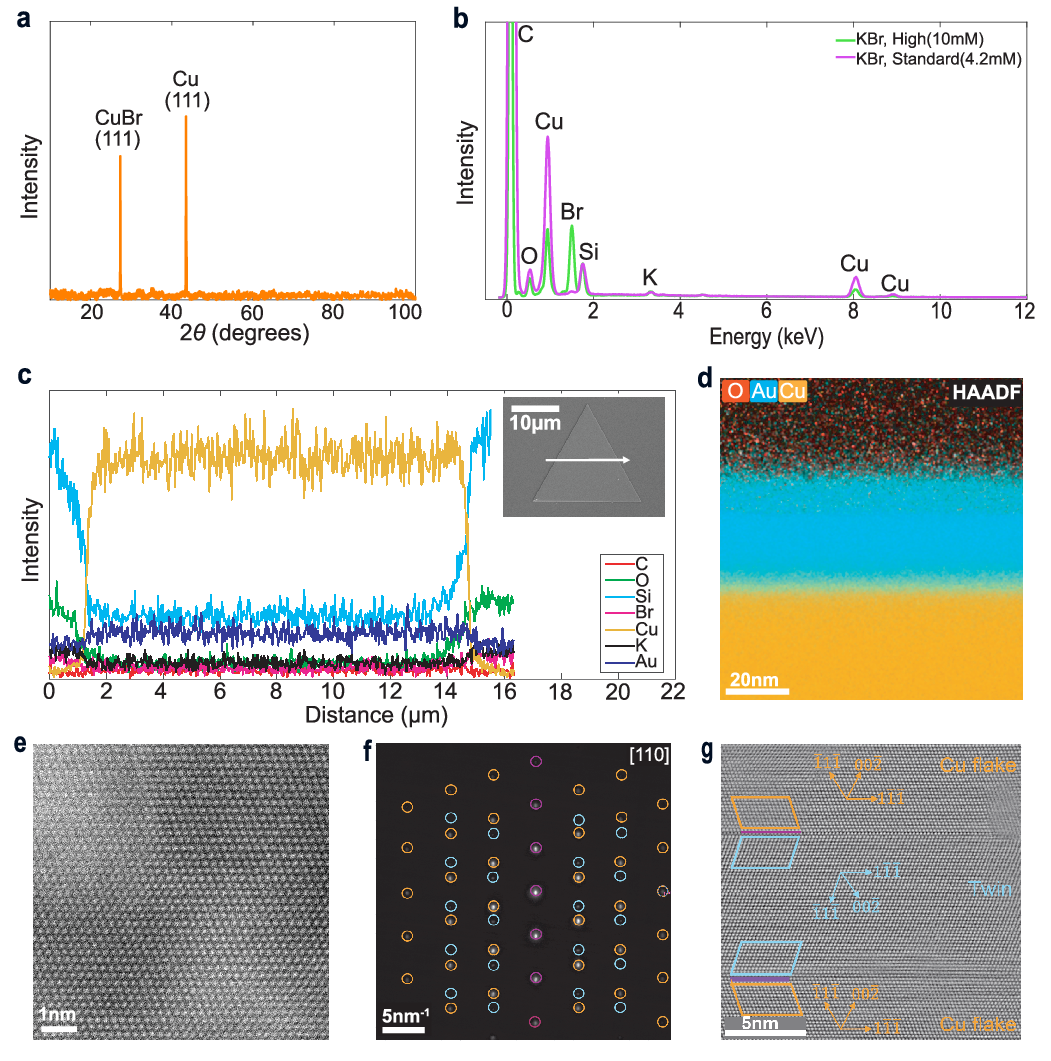}
    \caption{\justifying{ \textbf{Compositional analysis of the Cu flakes.} \textbf{a} XRD pattern of Cu flakes grown on a glass substrate. \textbf{b} EDS spectra taken from an area inside flake region for two samples prepared using different KBr concentrations of high (pink) and standard (green). The samples were coated with C for electrical conductivity. Si and O signal is introduced by the glass substrate underneath as seen in part (c).
    \textbf{c} EDS line scans for C(red), O(green), Si(cyan), Br(pink), Cu(orange), K(black) and Au(navy) recorded for the flake shown in the inset, along the direction indicated by the white arrow.The sample was coated with Au for electrical conductivity and to check for a carbon layer indicating an organic shell above the flake.  \textbf{d}  HAADF-STEM image of the cross-sectional Cu flake (orange) TEM lamella with conductive Au layer(blue) on top. The lack of O signal(red) on the flake region confirms the metallic nature of flake. \textbf{e} Atomically resolved HR-TEM image of the Cu flake. 
     \textbf{f} SAED pattern taken along the  [110] direction showing a sixfold symmetric fcc-Cu system along with distinct reflections from the nearby twinned region: orange circles mark reflections of the main Cu flake, blue circles mark reflections of the twin, and purple circles mark common reflections at the twin boundary. \textbf{g}  HR-TEM image of the twin region analyzed in (f).}}
    \label{fig-semtem}
\end{figure}

First, the X-Ray Diffraction (XRD) patterns hown in Figure \ref{fig-semtem}a  exhibits two prominent peaks: The strong diffraction peak at 43.5° confirms that Cu flakes are monocrystalline with \{111\} basal planes, while the smaller peak at 27.4° attributed to \{111\} CuBr is indicative of a bromide adlayer covering the basal surface\cite{fentahun_tunable_2021,lucas_cubr_2011}. It is interesting to note that the XRD measurement was performed one month after synthesis. Yet, no additional peaks corresponding to CuO or Cu$_2$O were observed, highlighting the pivotal role of the bromide ligand to mitigate surface oxidation and extend flake stability.  Moreover, the bright-light micrographs recorded on a flake right after it's been synthesized and three weeks under atmospheric conditions show the flakes to be preserved well (see Supplementary Figure 2), consistent with the XRD results.

 We observed that the Br signal from flakes grown with the standard recipe (4.2mM) was near the detection limit for EDS measurements. However, flakes with higher bromide content of 10mM showed a continuous bromide layer (Figure \ref{fig-semtem}b). Figure \ref{fig-semtem}c shows the elemental EDS line scans performed on the Cu flake included in the inset along the direction indicated by the arrow. A strong signal of Cu on the flake is visible, while the O signal comes from the glass substrate rather than the flake, further confirming the metallic nature of the flakes. Subsequently, we performed transmission electron microscopy (TEM) measurements on a cross-sectional Cu flake lamella prepared by focused ion beam-scanning electron microscopy (FIB-SEM, see Methods \ref{sec:Methods}). The insulating sample was sputter-coated with Au and C to improve the electrical conductivity. Figure \ref{fig-semtem}d shows the High-Angle Annular Dark-Field scanning transmission electron microscopy (HAADF-STEM) image in cross-section, where 0 (red) is not observed from the Cu region (orange).
We also coupled this measurement with electron energy loss spectroscopy (EELS) (see Supplementary Figure 4), where the O signal from the flake region is at noise levels. These analyses are in strong agreement with the SEM-EDS and XRD analyses and verifies the metallic nature of the Cu flakes.

Atomically resolved high-resolution TEM images (HRTEM) in Figure \ref{fig-semtem}e show the monocrystalline structure. The selective area diffraction (SAED) pattern in Figure \ref{fig-semtem}f was recorded from a nearby region along the [110] direction. The indexation performed with CrysTBox software unveils that the reflections match the fcc-Cu system \cite{klinger_crystallographic_2015}.The SAED pattern also reveals the presence of a twin boundary that extends parallel to the basal plane of the flake and mirrors the \{111\} lattice planes of the fcc.

The crucial role of stacking faults such as twin defects to sustain anisotropic growth of thin materials was introduced by Lofton and Sigmund and confirmed in several studies\cite{lofton_mechanisms_2005,golze_plasmon-mediated_2019, kiani_high_2022}.Twin planes provide side facets for the attachment of Cu adatoms and multiple twinning allows the side facets to regenerate each other. In addition, the energy barrier for addition of atoms in the vertical direction is increased, all together allowing realization of larger lateral sizes and higher aspect ratios. In accordance with this growth mechanism, we observe a twin defect along the flake that can favor the lateral growth, for which the SAED pattern and HRTEM images are shown in Figure \ref{fig-semtem}e,f respectively. Reflections from twinned regions are enclosed in blue, reflections from the Cu flake is in orange and lastly reflections from the twin mirror plane is marked in purple.

Lastly, absorptance measurements performed on the flakes were compared with the theoretical response of Cu of a similar thickness range (100-300nm) calculated by the transfer matrix method using previously reported optical paramaters for Cu \cite{mcpeak_plasmonic_2015}. The strong agreement between the experiment and theory further proves the metallic nature of the flakes (see Supplementary Figure 5).

\begin{figure}[ht!]
    \centering  
\includegraphics[width=17cm]{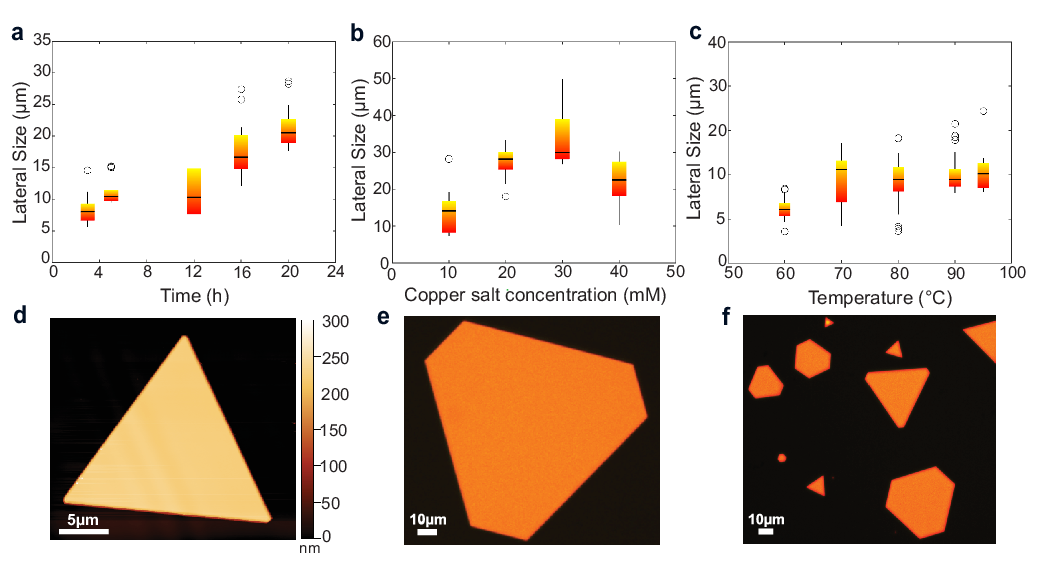}
\caption{\justifying{\textbf{Optimization of the flake size by parametric studies.} A single parameter was varied from the standard recipe at each time to acquire the trends shown in the boxplots. \textbf{a} Lateral size of the flake with respect to the growth period. \textbf{b} Influence of Cu sulfate precursor concentration on the lateral size, suggesting Cu ion depletion at lower concentrations of 0-20 mM and saturation above 30 mM. \textbf{c} Temperature dependence of the lateral size in the 60-95°C range, reaching a plateau near 90°C. \textbf{d} AFM image of a triangular flake synthesized using standard recipe in 20 hours with a uniform thickness of 225 nm along the flake.\textbf{e} Optical micrograph of a Cu flake grown using the optimized formulation of 20h reaction time, 95°C, 4.2 mM KBr, 30mM L-Ascorbic acid and 30 mM Cu salt precursor. The hexagonal shape confirms the \{111\} basal plane.\textbf{f} Optical micrograph showing increased yield of nanoparticles resulting from the recipe using optimized conditions.   }}    
\label{fig-growth}
\end{figure}

 Following the investigations on structural and chemical composition, we proceed to optimize the maximum lateral size by monitoring the effects of various factors such as the growth period, reaction temperature and salt precursor concentration on the growth mechanism. We start by varying the growth period between 3 and 20 hours, as shown in Figure \ref{fig-growth}a and observe that flakes larger than 5 $\mu$m are observed already after 3 hours and the average flake size increases over time, reaching a plateau around 20 hours. Based on the atomic force microscopy measurements we did not observe a direct correlation between the thickness of the flakes and the duration of growth (see Supplementary Figure 3). To understand whether reactions are limited by the initial availability of Cu ions, we changed the concentration of $\text{CuSO}_4 \cdot 5\text{H}_2\text{O}$ between 10-40 mM. Figure \ref{fig-growth}b shows that the largest flake sizes can be seen at 30 mM precursor concentration. We assume that the reaction is limited by the availability of Cu ions at low concentrations of 10-20 mM, while above 30 mM the solution is saturated and more by-products start to form.

At lower temperatures (60°C), growth predominantly occurs in the nucleation phase, producing smaller flakes, indicating that the reaction is constrained by limited kinetic energy. As shown in Figure \ref{fig-growth}c, the average lateral flake size increases at 70°C, suggesting accelerated growth; however, a broad size distribution persists due to the presence of numerous small flakes. With further elevation of the temperature, the size dispersion decreases, reflecting a more uniform growth, and the average size reaches a plateau with a few exceptionally large outliers. Since we use a water bath and deionized water as reaction medium, we remain at temperatures below 100°C to avoid boiling of the growth solution. Alternative growth media such as ethylene glycol were tried, yet no noticeable formation of flakes was seen, which may be linked to lower ion mobility in this medium.

Figure \ref{fig-growth}d shows an AFM image of a flake grown for 20 hours using the standard recipe that we have taken as a basis to perform the parametric studies. The thickness was found to be uniform (225 nm) across the entire flake surface, with a roughness mean factor of $\simeq$ 180 pm, indicating a smooth top surface. Overall, variation of a single parameter at a time formed flakes with lateral sizes in the range of 5 to 40$\mu$m, as seen from Figure \ref{fig-growth}a-d. The key improvement to lateral dimensions was observed by combining the optimal conditions of the parametric studies discussed to this point, i.e. 20 hours growth time, reaction temperature of 95°C, precursor concentration of 30 mM $\text{CuSO}_4 \cdot 5\text{H}_2\text{O}$ and 4.2 mM KBr.  We have succeeded in developing a formulation that produces Cu microflakes with lateral size of more than 130 $\mu$m, as shown in Figure \ref{fig-growth}e. Furthermore, this recipe allows improved yield of Cu flakes with few to no undesired by-products, as seen in Figure \ref{fig-growth}f. 

 It is worth noting that while most of the flakes still exhibited a flat top surface, we noticed for the first time formation of some flakes with steps on their top plane under these elevated conditions (see Supplementary Figure 6). It is possible to avoid these features without significantly compromising on lateral size by reducing the temperature to 90°C while maintaining the remaining parameters. However, flakes with these features can also offer an intriguing platform for energy conversion studies, as step-edge sites have been proven to be highly active catalytic sites that can reform mechanistic reaction pathways\cite{hendriksen_role_2010, zijlstra_vital_2020, rostrup-nielsen_step_2006}. These new exciting properties are unlocked thanks to our tailored surfactant-free on-substrate growth approach.

\section{Conclusions} \label{sec:Conclusions}
In this study, we report a highly repeatable, surfactant-free approach to synthesize large-area Cu microflakes on borosilicate glass substrates. By precisely tuning the halide type and concentration, we explore the powerful capabilities of bromide ions to control the flake morphology and enhance flake stability by mitigating surface oxidation. We then optimize the growth recipe by tuning reaction parameters such as growth time, temperature, and precursor concentrations. We achieve high-yield anisotropic growth, with an aspect ratio of up to 400 and a lateral size of more than 130 $\mu$m, making the maximum flake size two orders of magnitude higher than in previous studies. We perform various compositional analysis techniques such as SEM-EDS, STEM-EDS, EELS and XRD, which reveal the monocrystalline, fcc metallic nature of the Cu flakes. Copper is the most widely studied material for catalysis and energy conversion devices, thanks to its unique ability to support the formation of multi-carbon products on its surface. These large-area, surfactant-free Cu flakes with a well-defined (111) top surface provide a promising platform for catalysis, to perform controlled studies on selectivity, activity and complex reaction mechanisms. In addition, we envision these flakes to be highly beneficial for plasmonic and nanophotonic devices thanks to their single crystallinity.

\section{Methods} \label{sec:Methods}

\textbf{Synthesis and Characterization of Cu Microflakes} 
An experimental setup previously presented by Kiani and coworkers \cite{kiani_high_2022} was used. All chemical reagents were of analytical grade and were used as purchased from Sigma-Aldrich. First, certain concentrations of the salt precursor $\text{CuSO}_4 \cdot 5\text{H}_2\text{O}$ and L-Ascorbic acid(10mM of $\text{CuSO}_4 \cdot 5\text{H}_2\text{O}$ and 30 mM L-Ascorbic acid for the standard recipe) were added to 20 ml of ultrapure water in a standard 50-ml polypropylene falcon tube used as a reactor . Depending on the experiment, halides such as KBr, KI or KCl were added to the same solution in various concentrations (4.2 mM for the standard recipe) to serve as a capping and structuring agent. The final aqueous solution was then stirred vigorously for 30-60 minutes at room temperature.

To prepare the substrate, two borosilicate glass substrates (24 mm x 24 mm, $\#$1.5, Epredia) were cleaned by sonication in acetone and deionized water for 10 minutes each and dried by nitrogen blowing. The clean substrates were individually immersed in the aqueous growth solution, which was previously stirred homogeneously. The stirring was then stopped and substrates were placed in the tube with an upward tilted orientation so that they were immobilized during growth. The tube was then sealed and placed in a beaker filled with water and covered with foil to avoid possible light-induced effects. The system was then heated to the desired temperature (80°C, unless otherwise specified) at a ramp rate of 2°C/min on a hotplate without stirring. At the end of the growth window, heating was stopped and the substrates were immediately and individually removed to be thoroughly rinsed with ethanol and deionized water, three times each, followed by drying with nitrogen blowing in order to remove all unreacted species and reactants. The samples were stored either in an inert atmosphere or under ambient conditions for stability tests. 
\\
\\

\textbf{Materials Characterization} \\

Optical micrographs were recorded with a Nikon inverted optical microscope (Nikon Ti2A) and analyzed with the commercial software ImageJ (Ver.1.8.0). The X-Ray Diffraction experiment of regular theta-theta scans were performed on a Panalytical Empyrean X-Ray polycrystalline diffractometer in Bragg-Brentano geometry, equipped with long-focused sealed Cu X-Ray tube ($\lambda K\alpha = 1.5418 \, \text{\AA}$) and PIXcel 1D X-Ray detector. The patterns were collected in continuous mode between 10 and 100 degrees ($2\theta)$, with the step-size of 0.02626 degree over 19 hours. Background subtraction and peak identification were performed with HighScore plus v4.9 and PDF5+ v2024 \cite{degen_highscore_2014}. Absorptance measurements were performed with a configuration previously reported by Bowman et al\cite{bowman_best_2024}, on a customized system from NT\&C based on an inverted optical microscope (Nikon Ti2A). Scanning electron microscopic energy dispersive X‐ray spectroscopy (SEM-EDS) was performed using a ZEISS Merlin field emission microscope with an accelerating voltage of 15 kV and a working distance of 8.5 mm. Before the measurements, the insulating sample was coated with a Au (15nm) or C (12nm) layer by sputtering to achieve electrical conductivity. Atomic force microscopy (AFM) images and height profiles were acquired using a Bruker Fast-Scan AFM in ScanAsyst mode and the data was processed with Gwyddion commercial software. TEM cross-sectional analyzes were performed on a double aberration corrected FEI Titan Themis at 300 kV in [110] crystal direction of the Cu flake. The TEM lamella was fabricated from a Cu flake grown on the substrate with a dual beam focused ion beam (FIB)/SEM (Zeiss NVision 40). Prior to FIB milling, the sample was coated with a thin (ca. 20 nm) Au layer to ensure electrical conductivity, and the region of interest was finally protected by FIB-assisted carbon deposition (ca. 1 $\mu$m). The prepared fused silica/Cu(111)/Au/C stacked structure was extracted by FIB milling (30 kV Ga+ beam) and fixed on a Mo-TEM grid and stored under vacuum before measurements to prevent oxidation.  \\
\section*{Data Availability Statement} \label{sec:data}
The data underlying this manuscript will be available at Zenodo. 
\section*{Acknowledgements} \label{sec:acknowledgements}
 G.T., E.N.D., P.V. and F.K. acknowledge support of SNSF Eccellenza  Grant PCEGP2-194181. E.N.D and G.T. acknowledge Discovery Grant. D.P. acknowledges the support of Laidlaw Internship Program. A.R.B. acknowledges support of SNSF Eccellenza Grant PCEGP2-194181 and SNSF Swiss Postdoctoral Fellowship TMPFP2\_217040.  The authors also acknowledge the support of the following experimental facilities at EPFL: Centre for MicroNanofabrication (CMi) and EPFL: Interdisciplinary Centre for Electron Microscopy (CIME). We thank Milad Sabzehparvar for his assistance with carbon deposition. Finally, the authors would like to thank Dr. Wen Hua Bi for the help with XRD measurements, Dr. Barbara Bartova for TEM lamella preparation with FIB-SEM and Dr. David Reyes for the acquisition of the TEM images and data interpretation.

\section*{Author contributions} \label{sec:contributions}
G.T supervised all aspects of the project. E.N.D., D.P and P.V. carried out the synthesis and imaging of samples. E.N.D. performed data analysis, coordinated all the sample characterization. E.N.D. and P.V. carried out and analyzed AFM measurements. E.N.D., A.R.B. and O.C.K. performed and interpreted absorptance measurements. F.K. assisted SEM-EDX measurements and participated in data interpretation. E.N.D. and G.T. wrote the paper with input from all the authors.

\section*{Competing interests} \label{sec:comp}
The Authors declare no competing interests.
\section*{Figure Captions} \label{sec:figures}

\noindent \textbf{Figure 1: Halide assisted growth approach} \textbf{a} Illustration of the Cu flakes grown on substrate, planar (left) and cross-section (right) view. The halide adlayer (purple) promotes growth in the lateral direction. \textbf{b} Average lateral size obtained via using different halide ions. Resulting morphology for each condition are shown in c-f, with an exception for the use of KI, for which there was no visible formation of flakes. All scale bars indicate 5 $\mu$m. \textbf{c} Dependence of flake lateral size on KBr concentration. \textbf{d}  Anisotropic flakes with surface impurities were observed when no halide ions are used.  \textbf{e} Hexagonal flake obtained when using 4.2 mM of KBr.\textbf{f}  Etched flakes when KBr and KI are included simultaneously at equal concentrations of 4.2 mM.\textbf{g} Square-like flakes formed when 4.2 mM of KCl is used.
\\

\noindent  \textbf{Compositional analysis of the Cu flakes.} \textbf{a} XRD pattern of Cu flakes grown on a glass substrate. \textbf{b} EDS spectra taken from an area inside flake region for two samples prepared using different KBr concentrations of high (pink) and standard (green). The samples were coated with C for electrical conductivity. Si and O signal is introduced by the glass substrate underneath as seen in part (c).
    \textbf{c} EDS line scans for C(red), O(green), Si(cyan), Br(pink), Cu(orange), K(black) and Au(navy) recorded for the flake shown in the inset, along the direction indicated by the white arrow.The sample was coated with Au for electrical conductivity and to check for a carbon layer indicating an organic shell above the flake.  \textbf{d}  HAADF-STEM image of the cross-sectional Cu flake (orange) TEM lamella with conductive Au layer(blue) on top. The lack of O signal(red) on the flake region confirms the metallic nature of flake. \textbf{e} Atomically resolved HR-TEM image of the Cu flake. 
     \textbf{f} SAED pattern taken along the  [110] direction showing a sixfold symmetric fcc-Cu system along with distinct reflections from the nearby twinned region: orange circles mark reflections of the main Cu flake, blue circles mark reflections of the twin, and purple circles mark common reflections at the twin boundary. \textbf{g}  HR-TEM image of the twin region analyzed in (f).\\

\noindent \textbf{Optimization of the flake size by parametric studies.} A single parameter was varied from the standard recipe at each time to acquire the trends shown in the boxplots. \textbf{a} Lateral size of the flake with respect to the growth period. \textbf{b} Influence of Cu sulfate precursor concentration on the lateral size, suggesting Cu ion depletion at lower concentrations of 0-20 mM and saturation above 30 mM. \textbf{c} Temperature dependence of the lateral size in the 60-95°C range, reaching a plateau near 90°C. \textbf{d} AFM image of a triangular flake synthesized using standard recipe in 20 hours with a uniform thickness of 225 nm along the flake.\textbf{e} Optical micrograph of a Cu flake grown using the optimized formulation of 20h reaction time, 95°C, 4.2 mM KBr, 30mM L-Ascorbic acid and 30 mM Cu salt precursor. The hexagonal shape confirms the \{111\} basal plane.\textbf{f} Optical micrograph showing increased yield of nanoparticles resulting from the recipe using optimized conditions.   \\
\\


\bibliographystyle{unsrt}
\bibliography{main.bib} 

\begin{thebibliography}{10}

\bibitem{huang_freestanding_2011}
Xiaoqing Huang, Shaoheng Tang, Xiaoliang Mu, Yan Dai, Guangxu Chen, Zhiyou Zhou, Fangxiong Ruan, Zhilin Yang, and Nanfeng Zheng.
\newblock Freestanding palladium nanosheets with plasmonic and catalytic properties.
\newblock {\em Nature Nanotechnology}, 6(1):28--32, January 2011.
\newblock Publisher: Nature Publishing Group.

\bibitem{xiong_kinetically_2005}
Yujie Xiong, Joseph~M. McLellan, Jingyi Chen, Yadong Yin, Zhi-Yuan Li, and Younan Xia.
\newblock Kinetically {Controlled} {Synthesis} of {Triangular} and {Hexagonal} {Nanoplates} of {Palladium} and {Their} {SPR}/{SERS} {Properties}.
\newblock {\em Journal of the American Chemical Society}, 127(48):17118--17127, December 2005.
\newblock Publisher: American Chemical Society.

\bibitem{gao_highly_2012}
Chuanbo Gao, Zhenda Lu, Ying Liu, Qiao Zhang, Miaofang Chi, Quan Cheng, and Yadong Yin.
\newblock Highly stable silver nanoplates for surface plasmon resonance biosensing.
\newblock {\em Angewandte Chemie (International Ed. in English)}, 51(23):5629--5633, June 2012.

\bibitem{karaman_ultrafast_2024}
Can~O. Karaman, Anton~Yu Bykov, Fatemeh Kiani, Giulia Tagliabue, and Anatoly~V. Zayats.
\newblock Ultrafast hot-carrier dynamics in ultrathin monocrystalline gold.
\newblock {\em Nature Communications}, 15(1):703, 2024.
\newblock Publisher: Nature Publishing Group UK London.

\bibitem{pan_large_2024}
Chenxinyu Pan, Yuanbiao Tong, Haoliang Qian, Alexey~V. Krasavin, Jialin Li, Jiajie Zhu, Yiyun Zhang, Bowen Cui, Zhiyong Li, Chenming Wu, Lufang Liu, Linjun Li, Xin Guo, Anatoly~V. Zayats, Limin Tong, and Pan Wang.
\newblock Large area single crystal gold of single nanometer thickness for nanophotonics.
\newblock {\em Nature Communications}, 15(1):2840, April 2024.
\newblock Publisher: Nature Publishing Group.

\bibitem{sheng_copper_2022}
Aaron Sheng, Saurabh Khuje, Jian Yu, Thomas Parker, Jeng-Yuan Tsai, Lu~An, Yulong Huang, Zheng Li, Cheng-Gang Zhuang, Lanrik Kester, Qimin Yan, and Shenqiang Ren.
\newblock Copper {Nanoplates} for {Printing} {Flexible} {High}-{Temperature} {Conductors}.
\newblock {\em ACS Applied Nano Materials}, 5(3):4028--4037, March 2022.
\newblock Publisher: American Chemical Society.

\bibitem{kiani_interfacial_2023}
Fatemeh Kiani, Alan~R. Bowman, Milad Sabzehparvar, Can~O. Karaman, Ravishankar Sundararaman, and Giulia Tagliabue.
\newblock Interfacial {Hot} {Carrier} {Collection} {Controls} {Plasmonic} {Chemistry}, July 2023.

\bibitem{luc_two-dimensional_2019}
Wesley Luc, Xianbiao Fu, Jianjian Shi, Jing-Jing Lv, Matthew Jouny, Byung~Hee Ko, Yaobin Xu, Qing Tu, Xiaobing Hu, Jinsong Wu, Qin Yue, Yuanyue Liu, Feng Jiao, and Yijin Kang.
\newblock Two-dimensional copper nanosheets for electrochemical reduction of carbon monoxide to acetate.
\newblock {\em Nature Catalysis}, 2(5):423--430, May 2019.
\newblock Publisher: Nature Publishing Group.

\bibitem{yang_synthesis_2017}
Nailiang Yang, Zhicheng Zhang, Bo~Chen, Ying Huang, Junze Chen, Zhuangchai Lai, Ye~Chen, Melinda Sindoro, An-Liang Wang, Hongfei Cheng, Zhanxi Fan, Xiaozhi Liu, Bing Li, Yun Zong, Lin Gu, and Hua Zhang.
\newblock Synthesis of {Ultrathin} {PdCu} {Alloy} {Nanosheets} {Used} as a {Highly} {Efficient} {Electrocatalyst} for {Formic} {Acid} {Oxidation}.
\newblock {\em Advanced Materials}, 29(29):1700769, 2017.

\bibitem{roberts_high_2015}
F.~Sloan Roberts, Kendra~P. Kuhl, and Anders Nilsson.
\newblock High {Selectivity} for {Ethylene} from {Carbon} {Dioxide} {Reduction} over {Copper} {Nanocube} {Electrocatalysts}.
\newblock {\em Angewandte Chemie}, 127(17):5268--5271, 2015.

\bibitem{kumar_photochemical_2012}
Bhupendra Kumar, Mark Llorente, Jesse Froehlich, Tram Dang, Aaron Sathrum, and Clifford~P. Kubiak.
\newblock Photochemical and {Photoelectrochemical} {Reduction} of {CO2}.
\newblock {\em Annual Review of Physical Chemistry}, 63(Volume 63, 2012):541--569, May 2012.
\newblock Publisher: Annual Reviews.

\bibitem{zhao_charting_2022}
Qing Zhao, John Mark~P. Martirez, and Emily~A. Carter.
\newblock Charting {C}–{C} coupling pathways in electrochemical {CO} $_{\textrm{2}}$ reduction on {Cu}(111) using embedded correlated wavefunction theory.
\newblock {\em Proceedings of the National Academy of Sciences}, 119(44):e2202931119, November 2022.

\bibitem{dai_ultrastable_2017}
Lei Dai, Qing Qin, Pei Wang, Xiaojing Zhao, Chengyi Hu, Pengxin Liu, Ruixuan Qin, Mei Chen, Daohui Ou, Chaofa Xu, Shiguang Mo, Binghui Wu, Gang Fu, Peng Zhang, and Nanfeng Zheng.
\newblock Ultrastable atomic copper nanosheets for selective electrochemical reduction of carbon dioxide.
\newblock {\em Science Advances}, 3(9):e1701069, September 2017.
\newblock Publisher: American Association for the Advancement of Science.

\bibitem{li_thermal_2020}
Mo~Li, Alexandre Borsay, Mostapha Dakhchoune, Kun Zhao, Wen Luo, and Andreas Züttel.
\newblock Thermal stability of size-selected copper nanoparticles: {Effect} of size, support and {CO2} hydrogenation atmosphere.
\newblock {\em Applied Surface Science}, 510:145439, April 2020.

\bibitem{molahalli_properties_2024}
Vandana Molahalli, Aman Sharma, Kiran Bijapur, Gowri Soman, Apoorva Shetty, B.~Sirichandana, B.~G.~Maya Patel, Nattaporn Chattham, and Gurumurthy Hegde.
\newblock Properties, {Synthesis}, and {Characterization} of {Cu}-{Based} {Nanomaterials}.
\newblock In {\em Copper-{Based} {Nanomaterials} in {Organic} {Transformations}}, volume 1466 of {\em {ACS} {Symposium} {Series}}, pages 1--33. American Chemical Society, May 2024.
\newblock Section: 1.

\bibitem{wei_simple_2007}
Mingzhen Wei, Ning Lun, Xicheng Ma, and Shulin Wen.
\newblock A simple solvothermal reduction route to copper and cuprous oxide.
\newblock {\em Materials Letters}, 61(11):2147--2150, May 2007.

\bibitem{sreeju_microwave-assisted_2016}
N.~Sreeju., Alex Rufus, and Daizy Philip.
\newblock Microwave-assisted rapid synthesis of copper nanoparticles with exceptional stability and their multifaceted applications.
\newblock {\em Journal of Molecular Liquids}, 221:1008--1021, September 2016.

\bibitem{zhang_electrochemical_2014}
Q.~B. Zhang and Y.~X. Hua.
\newblock Electrochemical synthesis of copper nanoparticles using cuprous oxide as a precursor in choline chloride–urea deep eutectic solvent: nucleation and growth mechanism.
\newblock {\em Physical Chemistry Chemical Physics}, 16(48):27088--27095, November 2014.
\newblock Publisher: The Royal Society of Chemistry.

\bibitem{zaza_increasing_2024}
Ludovic Zaza, Dragos~C. Stoian, Noah Bussell, Petru~P. Albertini, Coline Boulanger, Jari Leemans, Krishna Kumar, Anna Loiudice, and Raffaella Buonsanti.
\newblock Increasing {Precursor} {Reactivity} {Enables} {Continuous} {Tunability} of {Copper} {Nanocrystals} from {Single}-{Crystalline} to {Twinned} and {Stacking} {Fault}-{Lined}.
\newblock {\em Journal of the American Chemical Society}, November 2024.
\newblock Publisher: American Chemical Society.

\bibitem{bowman_quantum-mechanical_2024}
Alan~R. Bowman, Alvaro Rodríguez~Echarri, Fatemeh Kiani, Fadil Iyikanat, Ted~V. Tsoulos, Joel~D. Cox, Ravishankar Sundararaman, F.~Javier García~de Abajo, and Giulia Tagliabue.
\newblock Quantum-mechanical effects in photoluminescence from thin crystalline gold films.
\newblock {\em Light: Science \& Applications}, 13(1):91, April 2024.
\newblock Publisher: Nature Publishing Group.

\bibitem{wang_sub-1_2023}
Ping Wang, Senyao Meng, Botao Zhang, Miao He, Pangen Li, Cheng Yang, Ge~Li, and Zhenxing Li.
\newblock Sub-1 nm {Cu2O} {Nanosheets} for the {Electrochemical} {CO2} {Reduction} and {Valence} {State}–{Activity} {Relationship}.
\newblock {\em Journal of the American Chemical Society}, 145(48):26133--26143, December 2023.
\newblock Publisher: American Chemical Society.

\bibitem{tang_role_2017}
Zengmin Tang, Hyunguk Kwon, Minyoung Yi, Kyungpil Kim, Jeong~Woo Han, Woo-Sik Kim, and Taekyung Yu.
\newblock Role of {Halide} {Ions} for {Controlling} {Morphology} of {Copper} {Nanocrystals} in {Aqueous} {Solution}.
\newblock {\em ChemistrySelect}, 2(17):4655--4661, 2017.

\bibitem{fu_alternative_2020}
Xianbiao Fu, Xingang Zhao, Xiaobing Hu, Kun He, Yanan Yu, Tao Li, Qing Tu, Xin Qian, Qin Yue, Michael~R. Wasielewski, and Yijin Kang.
\newblock Alternative route for electrochemical ammonia synthesis by reduction of nitrate on copper nanosheets.
\newblock {\em Applied Materials Today}, 19:100620, June 2020.

\bibitem{li_surfactant_2012}
Dongguo Li, Chao Wang, Dusan Tripkovic, Shouheng Sun, Nenad~M. Markovic, and Vojislav~R. Stamenkovic.
\newblock Surfactant {Removal} for {Colloidal} {Nanoparticles} from {Solution} {Synthesis}: {The} {Effect} on {Catalytic} {Performance}.
\newblock {\em ACS Catalysis}, 2(7):1358--1362, July 2012.
\newblock Publisher: American Chemical Society.

\bibitem{henglein_formation_2000}
Arnim Henglein.
\newblock Formation and {Absorption} {Spectrum} of {Copper} {Nanoparticles} from the {Radiolytic} {Reduction} of {Cu}({CN})2-.
\newblock {\em The Journal of Physical Chemistry B}, 104(6):1206--1211, February 2000.
\newblock Publisher: American Chemical Society.

\bibitem{wang_synthesis_2017}
Jing Wang, Xiaochuan Guo, Yan He, Mingjun Jiang, and Rong Sun.
\newblock The synthesis and tribological characteristics of triangular copper nanoplates as a grease additive.
\newblock {\em RSC Advances}, 7(64):40249--40254, August 2017.
\newblock Publisher: The Royal Society of Chemistry.

\bibitem{xu_preparation_2022}
Lijian Xu, Sijia Tang, Ling Zhang, Jingjing Du, Jianxiong Xu, Na~Li, and Zengmin Tang.
\newblock Preparation of {Copper} {Nanoplates} in {Aqueous} {Phase} and {Electrochemical} {Detection} of {Dopamine}.
\newblock {\em Life}, 12(7):999, July 2022.
\newblock Number: 7 Publisher: Multidisciplinary Digital Publishing Institute.

\bibitem{kiani_high_2022}
Fatemeh Kiani and Giulia Tagliabue.
\newblock High {Aspect} {Ratio} {Au} {Microflakes} via {Gap}-{Assisted} {Synthesis}.
\newblock {\em Chemistry of Materials}, 34(3):1278--1288, February 2022.
\newblock Publisher: American Chemical Society.

\bibitem{chen_reaction_2014}
Kunfeng Chen and Dongfeng Xue.
\newblock Reaction {Route} to the {Crystallization} of {Copper} {Oxides}.
\newblock {\em Applied Science and Convergence Technology}, 23(1):14--26, January 2014.

\bibitem{xia_shape-controlled_2009}
Younan Xia, Yujie Xiong, Byungkwon Lim, and Sara~E. Skrabalak.
\newblock Shape-{Controlled} {Synthesis} of {Metal} {Nanocrystals}: {Simple} {Chemistry} {Meets} {Complex} {Physics}?
\newblock {\em Angewandte Chemie International Edition}, 48(1):60--103, 2009.

\bibitem{schneider_nih_2012}
Caroline~A. Schneider, Wayne~S. Rasband, and Kevin~W. Eliceiri.
\newblock {NIH} {Image} to {ImageJ}: 25 years of image analysis.
\newblock {\em Nature Methods}, 9(7):671--675, July 2012.
\newblock Publisher: Nature Publishing Group.

\bibitem{guo_facile_2006}
Zhirui Guo, Yu~Zhang, Yun DuanMu, Lina Xu, Shengli Xie, and Ning Gu.
\newblock Facile synthesis of micrometer-sized gold nanoplates through an aniline-assisted route in ethylene glycol solution.
\newblock {\em Colloids and Surfaces A: Physicochemical and Engineering Aspects}, 278(1):33--38, April 2006.

\bibitem{li_edge_2012}
Jian-Min Li, Long-Gui Dai, Xiao-Ping Wan, and Xian-Lin Zeng.
\newblock An “edge to edge” jigsaw-puzzle two-dimensional vapor-phase transport growth of high-quality large-area wurtzite-type {ZnO} (0001) nanohexagons.
\newblock {\em Applied Physics Letters}, 101(17):173105, October 2012.

\bibitem{magnussen_-situ_1996}
O.~M. Magnussen, B.~M. Ocko, J.~X. Wang, and R.~R. Adzic.
\newblock In-{Situ} {X}-ray {Diffraction} and {STM} {Studies} of {Bromide} {Adsorption} on {Au}(111) {Electrodes}.
\newblock {\em The Journal of Physical Chemistry}, 100(13):5500--5508, January 1996.

\bibitem{ghosh_many_2018}
Sandeep Ghosh and Liberato Manna.
\newblock The {Many} “{Facets}” of {Halide} {Ions} in the {Chemistry} of {Colloidal} {Inorganic} {Nanocrystals}.
\newblock {\em Chemical Reviews}, 118(16):7804--7864, August 2018.

\bibitem{chen_high-yield_2014}
Lei Chen, Fei Ji, Yong Xu, Liu He, Yifan Mi, Feng Bao, Baoquan Sun, Xiaohong Zhang, and Qiao Zhang.
\newblock High-{Yield} {Seedless} {Synthesis} of {Triangular} {Gold} {Nanoplates} through {Oxidative} {Etching}.
\newblock {\em Nano Letters}, 14(12):7201--7206, December 2014.
\newblock Publisher: American Chemical Society.

\bibitem{fentahun_tunable_2021}
Daniel~A. Fentahun, Alekha Tyagi, Sugandha Singh, Prerna Sinha, Amodini Mishra, Somnath Danayak, Rajesh Kumar, and Kamal~K. Kar.
\newblock Tunable optical and electrical properties of p-type {Cu2O} thin films.
\newblock {\em Journal of Materials Science: Materials in Electronics}, 32(8):11158--11172, April 2021.

\bibitem{lucas_cubr_2011}
Francis~Olabanji Lucas, A.~Cowley, S.~Daniels, and P.~J. McNally.
\newblock {CuBr} blue light emitting electroluminescent thin film devices.
\newblock {\em physica status solidi c}, 8(9):2919--2922, 2011.

\bibitem{klinger_crystallographic_2015}
M.~Klinger and A.~Jäger.
\newblock Crystallographic {Tool} {Box} ({CrysTBox}): automated tools for transmission electron microscopists and crystallographers.
\newblock {\em Journal of Applied Crystallography}, 48(6):2012--2018, December 2015.
\newblock Publisher: International Union of Crystallography.

\bibitem{lofton_mechanisms_2005}
C.~Lofton and W.~Sigmund.
\newblock Mechanisms {Controlling} {Crystal} {Habits} of {Gold} and {Silver} {Colloids}.
\newblock {\em Advanced Functional Materials}, 15(7):1197--1208, 2005.

\bibitem{golze_plasmon-mediated_2019}
Spencer~D. Golze, Robert~A. Hughes, Sergei Rouvimov, Robert~D. Neal, Trevor~B. Demille, and Svetlana Neretina.
\newblock Plasmon-{Mediated} {Synthesis} of {Periodic} {Arrays} of {Gold} {Nanoplates} {Using} {Substrate}-{Immobilized} {Seeds} {Lined} with {Planar} {Defects}.
\newblock {\em Nano Letters}, 19(8):5653--5660, August 2019.
\newblock Publisher: American Chemical Society.

\bibitem{mcpeak_plasmonic_2015}
Kevin~M. McPeak, Sriharsha~V. Jayanti, Stephan J.~P. Kress, Stefan Meyer, Stelio Iotti, Aurelio Rossinelli, and David~J. Norris.
\newblock Plasmonic {Films} {Can} {Easily} {Be} {Better}: {Rules} and {Recipes}.
\newblock {\em ACS Photonics}, 2(3):326--333, March 2015.

\bibitem{hendriksen_role_2010}
Bas L.~M. Hendriksen, Marcelo~D. Ackermann, Richard van Rijn, Dunja Stoltz, Ioana Popa, Olivier Balmes, Andrea Resta, Didier Wermeille, Roberto Felici, Salvador Ferrer, and Joost W.~M. Frenken.
\newblock The role of steps in surface catalysis and reaction oscillations.
\newblock {\em Nature Chemistry}, 2(9):730--734, September 2010.
\newblock Publisher: Nature Publishing Group.

\bibitem{zijlstra_vital_2020}
Bart Zijlstra, Robin J.~P. Broos, Wei Chen, G.~Leendert Bezemer, Ivo A.~W. Filot, and Emiel J.~M. Hensen.
\newblock The {Vital} {Role} of {Step}-{Edge} {Sites} for {Both} {CO} {Activation} and {Chain} {Growth} on {Cobalt} {Fischer}–{Tropsch} {Catalysts} {Revealed} through {First}-{Principles}-{Based} {Microkinetic} {Modeling} {Including} {Lateral} {Interactions}.
\newblock {\em ACS Catalysis}, 10(16):9376--9400, August 2020.
\newblock Publisher: American Chemical Society.

\bibitem{rostrup-nielsen_step_2006}
Jens Rostrup-Nielsen and Jens~K. Nørskov.
\newblock Step sites in syngas catalysis.
\newblock {\em Topics in Catalysis}, 40(1):45--48, November 2006.

\bibitem{degen_highscore_2014}
Thomas Degen, Mustapha Sadki, Egbert Bron, Uwe König, and Gwilherm Nénert.
\newblock The {HighScore} suite.
\newblock {\em Powder Diffraction}, 29(S2):S13--S18, December 2014.

\bibitem{bowman_best_2024}
A.~R. Bowman, J.~Ma, F.~Kiani, G.~García~Martínez, and G.~Tagliabue.
\newblock Best practices in measuring absorption at the macro- and microscale.
\newblock {\em APL Photonics}, 9(6):061101, June 2024.

\end{thebibliography}

\end{document}